\documentclass[twocolumn,aps,pra,showpacs,superscriptaddress]{revtex4}
\usepackage[bf]{subfigure}
\usepackage{amsmath,amssymb}
\usepackage{graphicx}
\usepackage{times,txfonts}
\usepackage{color}
\usepackage{multirow}
\usepackage{bbold}
\usepackage{color}
\usepackage{soul}

\newcommand{\tr}{\operatorname{Tr}}
\newcommand{\ket}[1]{|#1\rangle}
\newcommand{\bra}[1]{\langle#1|}
\newcommand{\proj}[1]{\ket{#1}\bra{#1}}
\newcommand{\braket}[2]{\bra{#1}#2\rangle}
\newcommand{\id}{\mathbb{I}}

\begin{document}

\title{Experimental Entanglement Activation from Discord in a Programmable Quantum Measurement}

\author{Gerardo Adesso}
\affiliation{School of Mathematical Sciences, University of Nottingham, University Park, Nottingham NG7 2RD, United Kingdom}
\author{Vincenzo D'Ambrosio}
\affiliation{Dipartimento di Fisica, Sapienza Universit\`{a} di Roma, Roma 00185, Italy}
\author{Eleonora Nagali}
\affiliation{Dipartimento di Fisica, Sapienza Universit\`{a} di Roma, Roma 00185, Italy}
\author{Marco Piani}
\affiliation{Institute for Quantum Computing and Department of Physics and Astronomy, University of Waterloo, Waterloo N2L 3G1, Canada}
\author{Fabio Sciarrino}
\affiliation{Dipartimento di Fisica, Sapienza Universit\`{a} di Roma, Roma 00185, Italy}



%
%
%

\begin{abstract}
{In quantum mechanics, observing is not a passive act. Consider a system of two quantum particles $A$ and $B$: if a measurement apparatus $M$ is used to make an observation on   $B$, the overall state of the system $AB$ will typically be altered. 
When this happens no matter which local measurement is performed, the two objects $A$ and $B$ are revealed to possess peculiar correlations known as quantum discord. 
Here we demonstrate experimentally that the very act of local observation gives rise to an {\it activation protocol} which converts discord into distillable entanglement, 
a stronger and more useful form of quantum correlations,
 between the apparatus $M$ and the composite system $AB$. We adopt a flexible two-photon setup to realize a three-qubit system $(A,B,M)$ with programmable degrees of initial correlations, measurement interaction, and characterization processes.  Our experiment demonstrates the fundamental mechanism underpinning the ubiquitous act of observing the quantum world, and establishes the potential of discord in entanglement generation.}
\end{abstract}

\maketitle

%

The revolution brought in by quantum mechanics has required a deep change in the way we think about physics and about nature itself. At the same time, it has led to the development of disruptive technologies and, in recent years, to the birth of quantum information processing~\cite{NC}. Of the many ways in which quantum physics differs from classical one, two are especially striking: the measurement process leads almost always to some disturbance, and non-classical correlations---including but not reducing to quantum entanglement~\cite{entanglement}---can exist between separate physical systems.

These two key signatures of departure from classicality 
are deeply connected. To appreciate this, let us briefly review the model of measurement depicted by von Neumann~\cite{vonneumann}.
A complete measurement on a quantum system $B$ in an orthonormal basis $\ket{n}_B$ can be realized by letting $B$ interact with a measurement apparatus $M$, initialized in a fiducial pure state $\ket{0}_M$, through a controlled unitary $V_{BM}$ such that $V_{BM} \ket{n}_B \ket{0}_M = \ket{n}_B\ket{n}_M$. The measurement would then be accomplished by a readout of $M$. Let us consider the more general situation where $B$ is only a part of a composite system, its complement being denoted by $A$, and let us indicate with $\chi_{A|B}$ the initial state of the $AB$ system. Just before the readout, the \emph{premeasurement state}  \begin{equation}\rho_{AB|M} = (\id_A\otimes V_{BM})(\chi_{A|B} \otimes \ket{0}\bra{0}_M) (\id_A\otimes V^\dagger_{BM})\end{equation} will typically display quantum entanglement \cite{entanglement} between the apparatus $M$ and the $AB$ system.
  It  is  natural to wonder: \emph{When is the creation of entanglement between locally measured systems and apparatus (un)avoidable?}

\begin{figure}[t]
\centering
\subfigure[]{
\includegraphics[width=6.5cm]{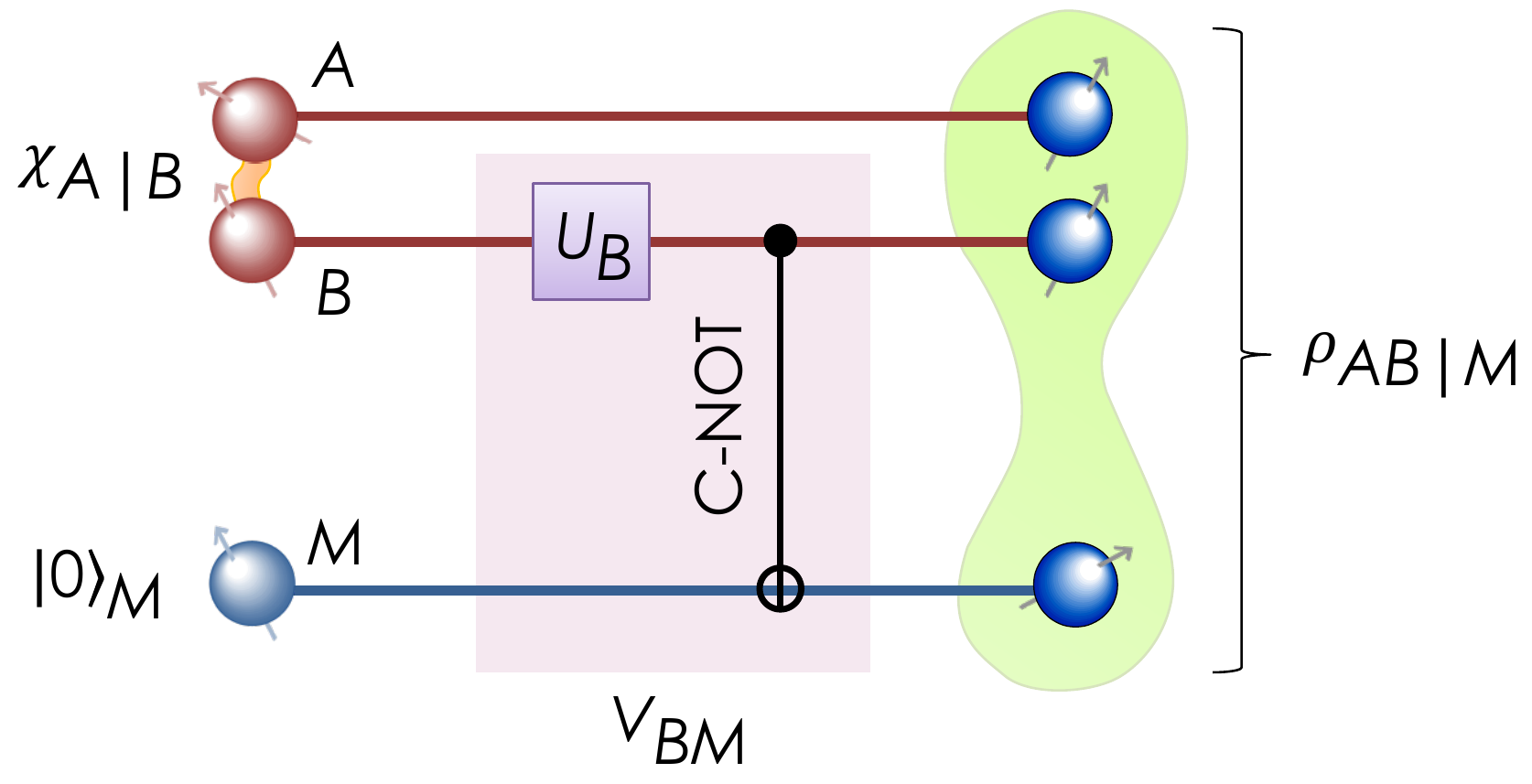}}
\subfigure[]{
\includegraphics[width=8cm]{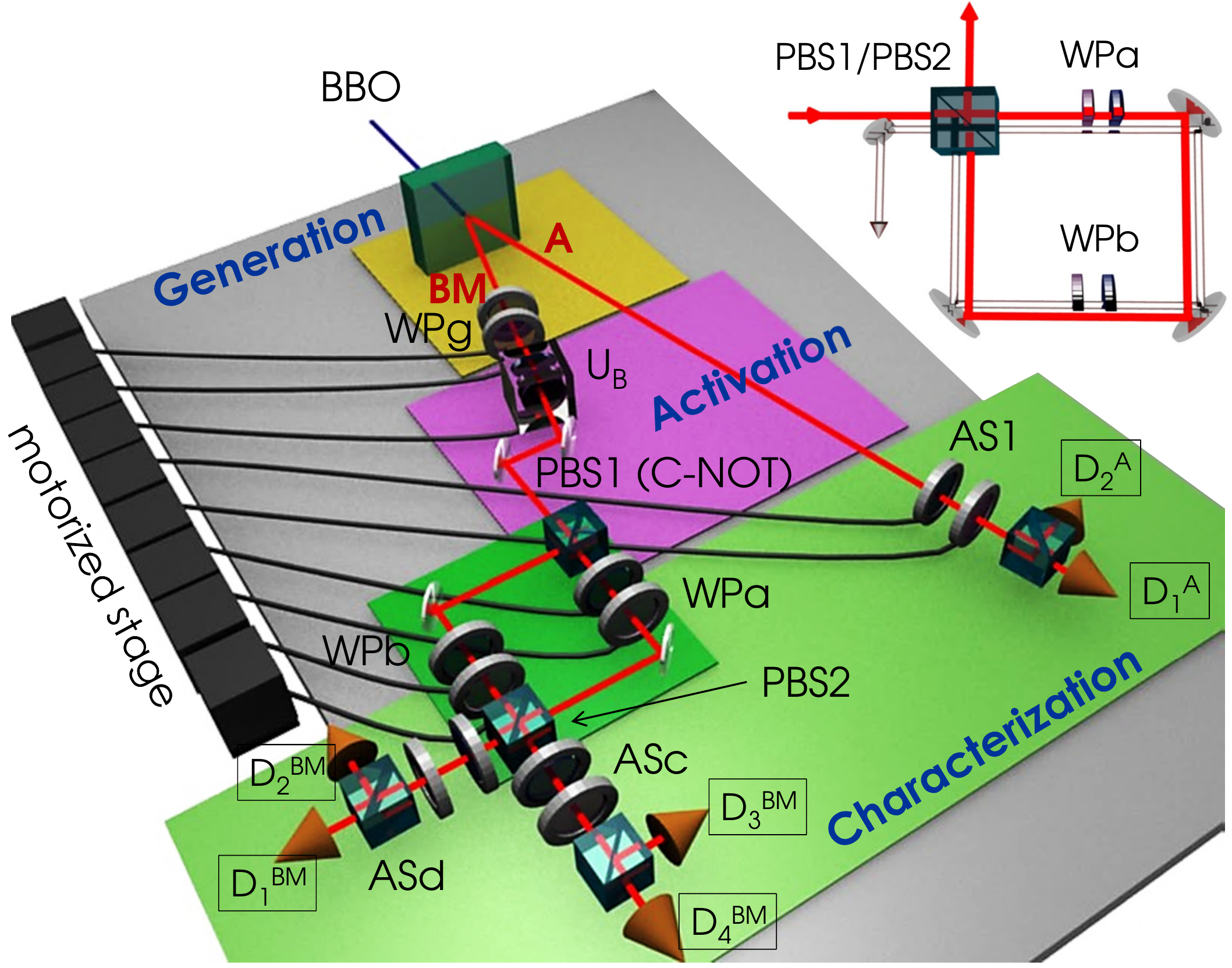}}\\
\caption{{\bf Activation protocol: design and experimental setup. 
}
(a) Circuit model for the activation protocol \cite{acti,streltsov,pianiadesso}. The initial discord of the state $\chi_{A|B}$ is activated into distillable entanglement for the state $\rho_{AB|M}$ by a local variable operation ($V_{BM}$) composed by a single qubit operation on $B$ ($U_{B}$) and a C-NOT between $B$ and $M$.
(b) Experimental setup. Two polarization entangled photons are generated in a beta barium borate crystal (BBO) via spontaneous parametric down conversion \cite{Kwia95}. The entangled state is changed in different Bell states by adopting  birefringent waveplates (WPg).
The $U_{B}$ is realized by two waveplates \cite{noteU} while the C-NOT is realized by the polarizing beam splitter (PBS1) which couples polarization and path of photon BM. To characterize the output state we adopt waveplates WPa, WPb and analysis stages ASc, ASd and AS1. Finally, single-photon detectors (${\text{D}_i}^{\text{A}}$, ${\text{D}_j}^{\text{BM}}$) measure photon coincidences. All the WPs are mounted on motorized stages controlled via computer. The inset displays the equivalent displaced Sagnac interferometric scheme adopted in the experiment \cite{Naga12,Walb06,Okam11}.
}
\label{setup}
\end{figure}

The answer is intertwined with the concept of {\it quantum discord} \cite{zurek,vedral}, a nonclassical signature in composite systems which corresponds to the amount of correlations between two or more parties necessarily destroyed during a minimally disturbing local measurement,
and similarly quantifies the minimal informational disturbance induced on the system by such a measurement
\cite{zurek}. Quantum discord, as a representative of a general type of nonclassical correlations \cite{reviewmodi}, has received widespread attention both for its fundamental role in defining the border between classical and quantum world \cite{zurekrev,acti,streltsov,lqu,newalex,marcodarwin}, and for its possible resource power for information processing and quantum computation, even in absence of entanglement \cite{datta,white,concordant,melo,cavalcantidiscord,npgu,npdakic,entdist1,entdist2,entdistexp,interpower,merali}. This power, however, has not been fully demonstrated to date \cite{horo,renzie,reviewmodi}.

In Ref.~\cite{acti,streltsov}, a general result was proven: there exists only a special class of states of $AB$ for which one can measure $B$ in some basis $\ket{n}_B$ without inducing disturbance, and such that no entanglement is generated between $M$ and $AB$ in the premeasurement stage. These  are the quantum-classical states with zero discord (from the perspective of $B$), taking the form $\varpi_{A|B} = \sum_n p_n \tau^n_A \otimes \ket{n}\bra{n}_B$, where $p_n$ is a probability distribution and each $\tau^n_A$ is a density operator for  $A$.  For any other state $\chi_{A|B}$, its amount of quantum discord determines  the minimum entanglement which is  {\it activated}, i.e., necessarily established with the apparatus $M$ during a local probing of  $B$. This was theoretically established in \cite{acti,streltsov}, leading to the proposal to quantify discord exactly in terms of the minimum activable entanglement (see also \cite{maurolaura,giovannetti} for alternative schemes). Depending on how the created entanglement is quantified, measuring discord via activation may coincide with other approaches \cite{reviewmodi}, e.g.~related to distances from the set of quantum-classical states \cite{modiunif,taka}.

The observation of the predicted  qualitative and quantitative correspondence between discord and generated entanglement is the main focus of this Letter.
Besides the foundational relevance, the fact that entanglement is necessarily generated during a measurement renders discord useful in schemes aimed at producing entanglement for quantum technological applications~\cite{acti,quantumnessIJQI,coleshierarchy}. It is worth mentioning that, once system-apparatus entanglement is  generated in the activation framework thanks to initial discord between subsystems $A$ and $B$, such entanglement can be used flexibly, and in particular be mapped into $A|B$ entanglement via local operations and classical communication~\cite{quantumnessIJQI}.

\begin{figure}[t!]
\centering
\subfigure[]{\includegraphics[width=4cm]{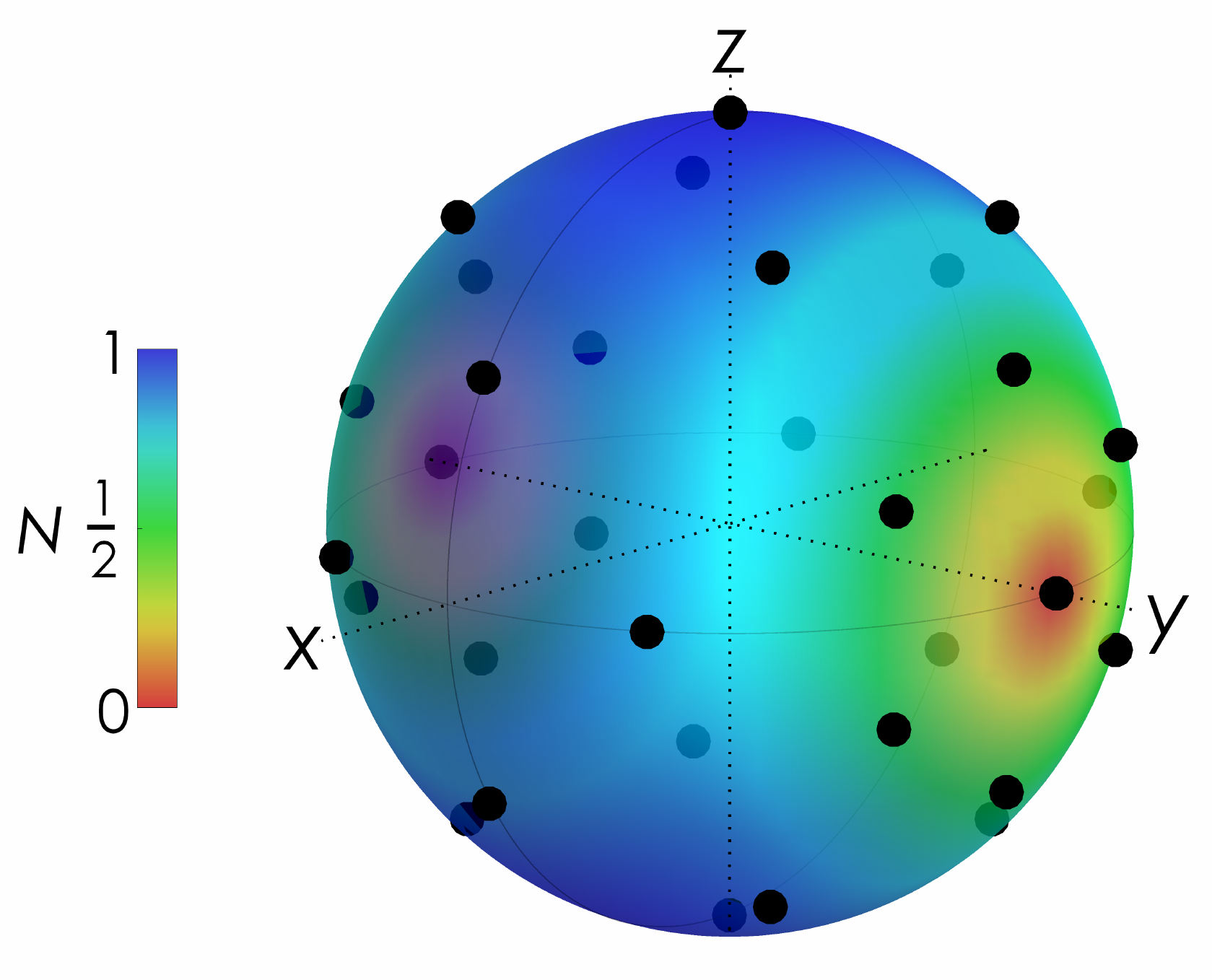}}
\subfigure[]{
\includegraphics[width=8.5cm]{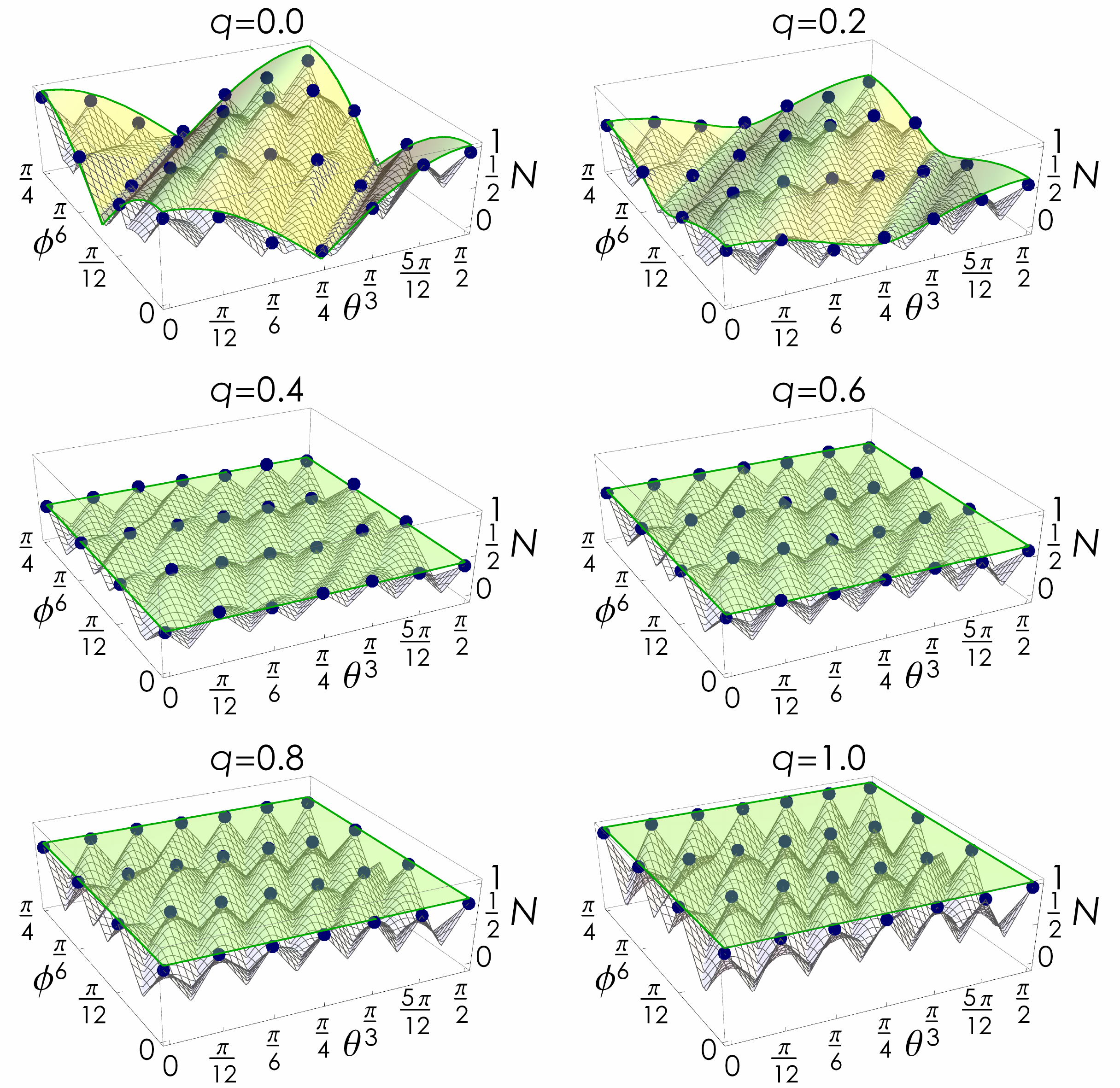}}
\caption{{\bf Activation protocol: experimental results.}
(a) Measurement bases for qubit $B$ defined by Bloch vectors $\pm\vec{n}(\theta_j,\phi_k)$ (dots), where $\theta,\phi$ are waveplates angles \cite{noteU}; the Bloch sphere is overlayed with a density plot of the expected negativity of the output premeasurement state $N(\rho^{(q,\theta,\phi)}_{AB|M})$ for $q=0$.
(b) Comparative results of the activation protocol. Each plot corresponds to a different value of $q$, and within each panel $28$ different settings are varied for $U_B$. Each blue point amounts to the entanglement $N(\rho_{AB|M}^{(q,\theta_j,\phi_k)})$ of the corresponding premeasurement state reconstructed by tomography; the errors, evaluated from the Poissonian statistics of photon events, are below $10^{-2}$. In each plot, we include the lower bound $N_{\rm low}(\rho_{AB|M}^{(q,\theta,\phi)})$ on the negativity derived in Eq.~(B.5) of \cite{epaps} (wireframe meshed surface), which remains positive for $q>0$ confirming that we sampled enough settings for $U_B$ to certify the unavoidable entanglement creation from initial discord. The green translucent smooth surfaces correspond to theoretical expectations  $N_{\rm th}(\rho_{AB|M}^{(q,\theta,\phi)})$ \cite{epaps}.
}
\label{fignets}
\end{figure}

We demonstrate experimentally the entanglement activation from discord by realizing a programmable quantum measurement process with bulk optics, see Figure~\ref{setup}. We consider a family of initial states $\chi_{A|B}$ of two qubits $AB$ with variable degree of discord, and we use a third logical qubit $M$ (as the apparatus) to perform a variety of local measurements on $B$. Taken {\it prima facie}, the verification of the results of \cite{acti,streltsov} would require to implement a continuous set of measurements, which is impossible in practice. We develop a rigorous procedure to define a discrete net of settings which reliably approximates a continuous sampling over the Bloch sphere of $B$, see Figure~\ref{fignets}.
This novel approach is reminiscent of the notion of $\epsilon$-net for metric spaces \cite{enet1,enet2} and tackles the problem of considering a worst case scenario over a continuous set, at variance with other experiments where only average values are cared for~\cite{twirling} or polynomial interpolation is invoked~\cite{lobino2008}.  Based on a finite set of data, we conclude that as soon as the initial state $\chi_{A|B}$ is nonclassically correlated, entanglement is activated for {\it any} possible measurement setting. The minimum amount of entanglement activated, over all possible local measurements, agrees quantitatively with the suitably quantified initial amount of discord, verifying the theoretical predictions \cite{acti,streltsov,quantumnessIJQI,pianiadesso}. Furthermore, when the initial state $\chi_{A|B}$ is itself entangled, genuine tripartite entanglement is detected among $A$, $B$, and $M$ in the premeasurement state $\rho_{AB|M}$ \cite{pianiadesso} by means of a witness operator \cite{witness}. Our work puts quantum discord on the firm ground of an operational ingredient for entanglement generation.

{\bf Activation protocol: description and implementation.}\\
The experimental implementation of the activation protocol is based on a two-qubit system $AB$ encoded in the polarization states of two photons ``A'' and ``BM'', and an ancillary qubit $M$ corresponding to the two possible paths that  photon BM  can take. The protocol is divided in three stages, see Figure~\ref{setup}.

{\bfseries \em Generation.}
By exploiting spontaneous parametric down conversion in a non-linear crystal, we generate two polarization maximally entangled photons in the Bell singlet state $\ket{\Psi^-}_{A|B}$, with $\ket{\Psi^\pm}_{A|B}=\frac{1}{\sqrt{2}}(\ket{H}_{A}\ket{V}_{B}  \pm \ket{V}_{A} \ket{H}_{B})$, where $(H, V)$ are linear horizontal and vertical polarization, respectively \cite{Kwia95}.
In order to complete the Bell basis, the three other Bell states $\ket{\Psi^+}_{A|B}$ and $\ket{\Phi^{\pm}}_{A|B}=\frac{1}{\sqrt{2}}(\ket{H}_{A}\ket{H}_{B} \pm\ket{V}_{A} \ket{V}_{B})$ are generated by applying to $\ket{\Psi^-}_{A|B}$ a suitable combination of local unitary Pauli operators $\{\sigma_i\}$, implemented by exploiting the effect of birefringent waveplates on photon BM (WPg in Figure \ref{setup}).
Thus a generic Bell diagonal mixed state $\chi_{A|B}$ can be obtained by switching between different Bell states for an appropriate time depending on the weight of the corresponding state in the statistical mixture \cite{marcobound}.

{\bfseries \em Activation.}
The activation procedure for the state $\chi_{A|B}$ consists of two steps. The interaction $V_{BM}$ is realized by first applying a variable unitary operation $U_B$ to system $B$, which plays the role of selecting the measurement basis $\ket{n}_B$, i.e., the observable to be measured. Up to irrelevant phases, the basis $\ket{n}_B$ is uniquely defined by a Bloch vector $\vec{n}$ \cite{noteU}.
Then $B$ interacts with the apparatus $M$ via a C-NOT gate, see Figure~\ref{setup}(a). After these local operations, the initial discord of the state $\chi_{A|B}$ is activated into distillable entanglement in the state $\rho_{AB|M}$.
Experimentally the operation $U_B$ on the polarization of photon BM is implemented by the two waveplates ${\mathrm{U_B}}$ in Figure~\ref{setup}; the C-NOT gate is achieved by exploiting a polarizing beam splitter (PBS1) which implements the following transformation: $\ket{H}_B\ket{a}_M \rightarrow \ket{H}_B\ket{a}_M$ and  $\ket{V}_B\ket{a}_M \rightarrow \ket{V}_B\ket{b}_M$, where the path states  $\{\ket{a}_M,\ket{b}_M\}$ define the computational basis for the logical qubit $M$.

{\bfseries \em Characterization.}
To measure the activated entanglement we fully reconstruct the output premeasurement state $\rho_{AB|M}$ via quantum state tomography \cite{Jame01}, requiring projective measurements over polarization of both photons and path of photon BM. While polarization measurement for photon A is easily performed with two waveplates  and a PBS (AS1), photon BM requires a more complex setup based on an interferometer which is shown in Figure~\ref{setup}. In each arm of the interferometer a quarter-waveplate and a half-waveplate are exploited to measure polarization of photon BM; depending on this measurement, the photon will leave PBS2 from exit $c$ or exit $d$. Then the polarization state of photon BM depends only on its path inside the interferometer, hence a polarization analysis stage (ASc and ASd) is sufficient to perform a measurement of qubit $M$. Instead of the  Mach-Zehnder interferometer of Figure~\ref{setup}, in the experiment we adopted a more compact and stable displaced Sagnac interferometer \cite{Naga12,Walb06,Okam11}, see inset of Figure~\ref{setup}(b). Both photons are finally collected by single-mode fibers and sent to single-photon detectors; see also \cite{epaps}.

The activation protocol does not need destructive measurements, hence the generated entanglement can be used for further processing. In our  experimental setup, the entanglement is activated after PBS1 which implements a C-NOT gate between polarization and path of photon BM. As an alternative approach one could encode qubits $A$, $B$ and $M$ in three different photons or other physical systems like trapped ions or superconducting qubits. The destructive measurements are performed here only in order to verify the activation process.

{\bf Experimental state preparation and characterization.}
Following the above procedure, we prepared the qubits $A$ and $B$ in a family of symmetric Bell diagonal states of the form
\begin{equation}\label{statoq}
\chi^{(q)}_{A|B}=q \ket{\Psi^+}\bra{\Psi^+}_{A|B} + \frac{1-q}{2} \left(
\ket{\Phi^+}\bra{\Phi^+}_{A|B}+\ket{\Psi^-}\bra{\Psi^-}_{A|B}\right)\,.
\end{equation}
Here the parameter $q \in [0,1]$ regulates the quantumness of correlations in the initial state $\chi^{(q)}_{A|B}$: for $q=0$ the state is only classically correlated, for any $q>0$ it has discord, while for $q>1/2$ the state is further entangled. The activation theorem predicts that for any $q>0$, the premeasurement state $\rho^{(q,U_B)}_{AB|M}$ is entangled for all possible choices of $U_B$. The minimum entanglement in the premeasurement state over all choices of $U_B$, measured e.g.~by the negativity $N$ \cite{entanglement,negativity}, defines a measure of discord for the initial state known as negativity of quantumness \cite{acti,taka,isabela}, $Q_N(\chi^{(q)}_{A|B})=\min_{U_B} N(\rho^{(q,U_B)}_{AB|M})$. When $B$ is a qubit, like in our case, the negativity of quantumness coincides with an independently defined geometric measure of discord $D(\chi_{A|B})$ given by the trace-distance  between $\chi_{A|B}$ and the closest quantum-classical state $\varpi_{A|B}$ \cite{taka,sarandy} (see Appendix). In particular, for the class of states in \eqref{statoq}, the trace-distance discord reads $D(\chi^{(q)}_{A|B})=q$ while their initial entanglement measured by the negativity  \cite{entanglement,negativity} is $N(\chi^{(q)}_{A|B}) = \max(0,2q-1)$.

In the experiment we pick $6$ evenly spaced representative values for $q$. The operator $U_B$ is implemented through a combination of a quarter-waveplate and a half-waveplate whose optical axes are rotated respectively by $\theta$ and $\phi$ with respect to the direction of linear horizontal polarization, which overall amounts to measuring $B$ in a Bloch direction $\pm \vec{n}(\theta,\phi)$ \cite{noteU}.
 We define a discrete net of values $(\theta_j,\phi_k)$,  which provides an adequate sampling---reminiscent of the notion of $\epsilon$-net for metric spaces \cite{enet1,enet2}---of the Bloch sphere for the  purpose of entanglement activation, see \cite{epaps} for the mathematical details. This procedure brings us down to measure $N(\rho^{(q,\theta_j,\phi_k)}_{AB|M})$ in $28$ different settings $(j,k)$ for any chosen value of $q$, defined by the waveplates angles $(\theta_j,\phi_k)$ with $\theta_j = j \frac{\pi}{12}, \phi_k=k \frac{\pi}{12}$, $j=0,\ldots,6$, $k=0,\ldots,3$. The corresponding Bloch directions $\pm \vec{n}(\theta_j,\phi_k)$ are displayed in  Figure~\ref{fignets}(a) \cite{noteU}. The waveplates angles are programmed by a computer-controlled motorized stage as illustrated in Figure~\ref{setup}(b). An example of tomographically reconstructed state is reported in \cite{epaps}.

{\bf Demonstration of entanglement activation.}
The activation results are shown in Figure~\ref{fignets}(b) together with the closely matching theoretical surface based on Eq.~(A.1) of \cite{epaps} which considers an ideal implementation of the state in Eq.~(\ref{statoq}). We observe a satisfactory agreement with the experimental acquisitions which confirms our high degree of control on all stages of the experiment.
Notice how the case $q=0$, where the initial state $\chi_{A|B}$ of Eq.~(\ref{statoq}) has zero discord, is the only case in which for certain values of $U_B$ the apparatus $M$ does not necessarily entangle with the observed system $AB$, in accordance with the theoretical formulation of the activation protocol \cite{acti,streltsov}.

We complete our plots with a grid of lower bounds to $N(\rho^{(q,\theta,\phi)}_{AB|M})$ for arbitrary $\theta,\phi$. The bounds, which originate from purposely derived continuity limits for the negativity \cite{epaps}, allow us to infer---based on the finite net of experimental data---that for $q>0$ the premeasurement state will always display entanglement between $M$ and the  $AB$ for all possible measurement choices on $B$, hence providing a sound verification of the activation theorem \cite{acti,streltsov}. Only in the case $q=0$, when the initial state $\chi_{A|B}$ of Eq.~(\ref{statoq}) has zero discord, we find  that for certain measurement settings $N(\rho^{(0,\theta_j,\phi_k)}_{AB|M})$ is essentially vanishing within the experimental imperfections. The output entanglement was generally found to be minimized by $\theta_j=\pi/4,\phi_k=0$; while for $q>1/3$, any choice of $U_B$ produces the same entanglement in the premeasurement state apart from experimental fluctuations, see Figure~\ref{fignets}(b).

\begin{figure}[t]
\subfigure[]{
\includegraphics[height=4cm]{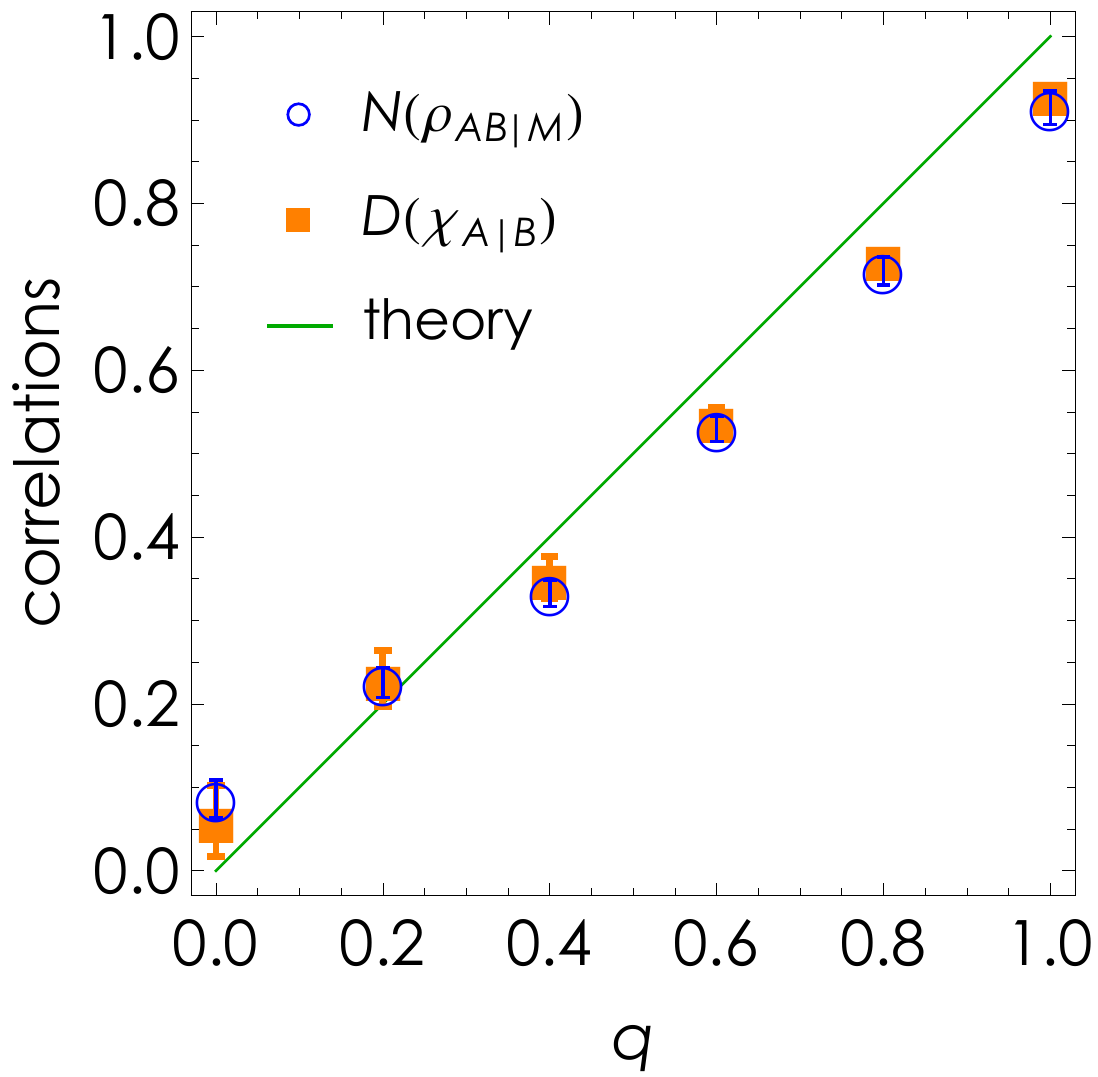}}\hspace*{.1cm}
\subfigure[]{
\includegraphics[height=4cm]{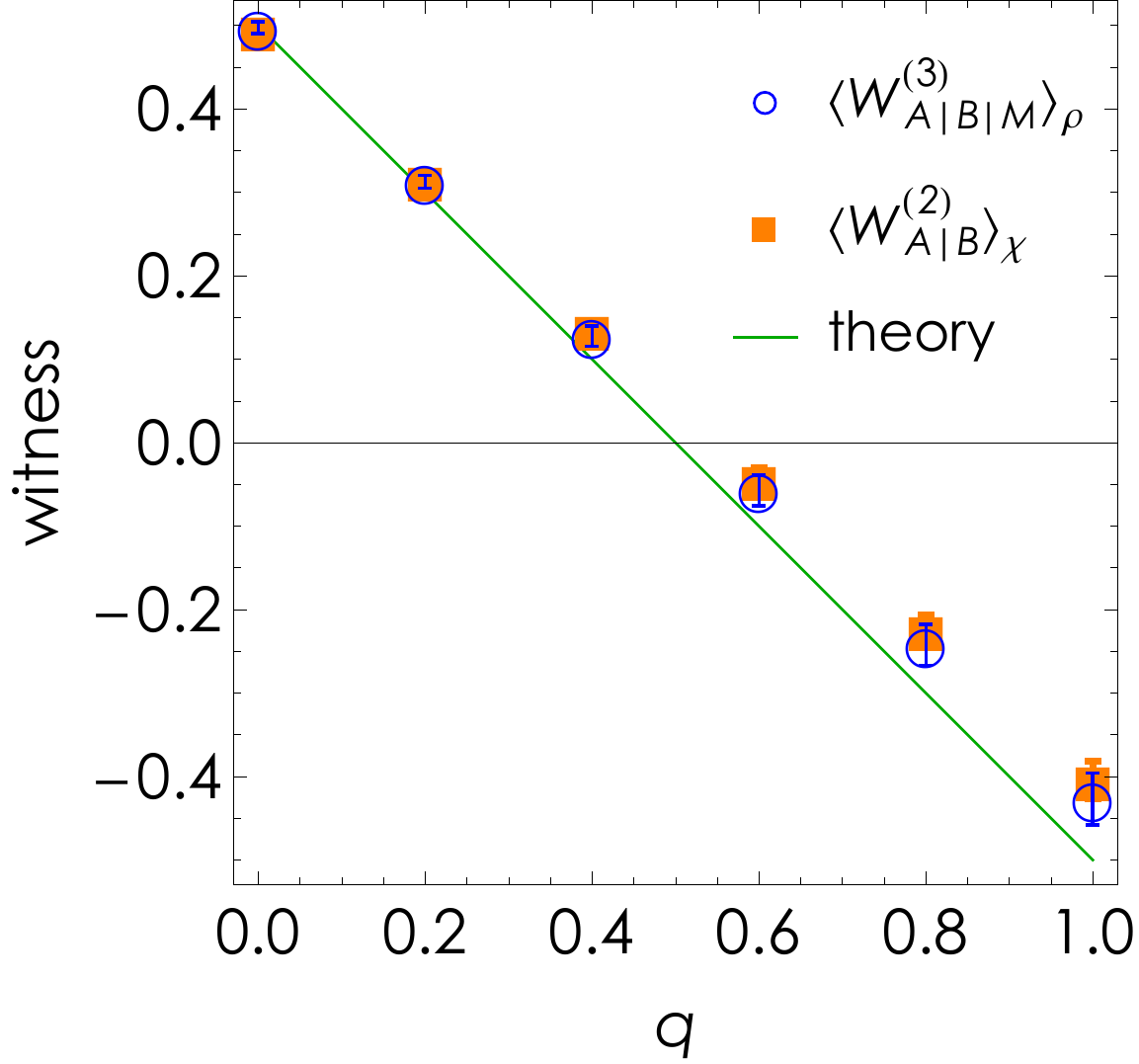}}
\caption{{\bf Activation protocol: quantitative verification.}
(a) 
The minimum output entanglement activated with the apparatus, measured by the negativity $\min_{(\theta,\phi)} N\big(\rho_{AB|M}^{(q,\theta,\phi)}\big)$, is verified to match the input discord $D(\chi_{A|B}^{(q)})$  in the system, measured by the trace-distance from the set of quantum-classical states.
(b) Input bipartite entanglement activates into genuine tripartite entanglement \cite{pianiadesso}. Measured expectation values of two witness operators (negative values detect entanglement) \cite{entanglement,witness} are plotted; $W^{(2)}_{A|B}$  is a witness for bipartite entanglement in the initial state $\chi^{(q)}_{A|B}$, while $W^{(3)}_{A|B|M}$  is a witness for genuine tripartite Greenberger-Horne-Zeilinger--type entanglement among $A$, $B$ and $M$ in the premeasurement state $\rho^{(q)}_{AB|M}$ \cite{epaps}.}
\label{figattiva}
\end{figure}

We can now exploit our data to verify the quantitative prediction of the activation theorem \cite{acti,streltsov}, as shown in Figure~\ref{figattiva}(a). We find experimentally that the minimum output entanglement $\min_{j,k}N(\rho^{(q,\theta_j,\phi_k)}_{AB|M})$ generated with the apparatus, measured by negativity, precisely matches the amount of trace-distance discord $D(\chi^{(q)}_{A|B})$ initially detected between $A$ and $B$ \cite{epaps}. This demonstrates the operational power of discord as  ``entanglement potential'' \cite{acti,streltsov,melo}.
 and  proves successfully the  activation of distillable bipartite entanglement from input discord.
Small discrepancies (of at most $0.1$ units) from  the theoretical green line can  be traced back to imperfections in the initial preparation of the four polarization Bell states of photons A and BM, which in our implementation reach a mean purity of $(93.6 \pm 0.2)\%$, consistent e.g.~with the experimental value for $D(\chi_{A|B}^{(q=1)})$.

Finally, in case the initial state $\chi^{(q)}_{A|B}$ is entangled as well, we observe that genuine tripartite entanglement is created in the premeasurement state among $A$, $B$, and $M$ as a result of the activation protocol \cite{pianiadesso}. This is verified by detecting suitable entanglement witnesses \cite{entanglement,witness} for input and output states \cite{epaps} as presented in Figure~\ref{figattiva}(b).
 Theoretically, the expectation values for both witnesses  on the corresponding states should coincide and be given by $\frac12-q$ (solid green line); we find a satisfactory agreement between the two datasets and the theory, detecting in particular negative expectation values for the two witnesses in the relevant parameter range $q>1/2$.

{\bf Conclusions.}
The experimental observation of the activation protocol, presented as an abstract theorem in \cite{acti,streltsov}, brings the notion of quantumness in composite systems to a more concrete level than ever before. Our demonstration establishes the usefulness of various forms of nonclassical correlations as resources for information processing: discord is necessarily activated into bipartite entanglement, and bipartite entanglement into genuine tripartite entanglement, by the  act of performing a local measurement.
We believe an approach based on an $\epsilon$-net construction such as ours can effectively be adopted in future experiments aimed at substantiating mathematical predictions via sampling continuously distributed parameters, in particular in worst case scenarios. The implementation of a von Neumann chain \cite{vonneumann,pianiadesso} may be further considered, exploiting recent advances in integrated quantum photonics \cite{Geremia,Cres11,Cres13}.
We hope our work can stimulate further endeavours towards a deeper physical understanding and practical exploitation of {\it quantumness}, in its manifold manifestations, for information processing and communication.

\textbf{Appendix. Measures of correlations.}
The negativity, a measure  of entanglement for a bipartite state $\chi_{A|B}$, can be defined as
$N(\chi_{A|B}) = \|\varrho_{A|B}^\Gamma\|_1-1$,
where the suffix $\Gamma$ denotes partial transposition \cite{entanglement}, and $\| O \|_1=\text{Tr}\sqrt{O^\dagger O}$ is the trace norm, i.e.~the sum of the singular values of $O$.
The negativity of quantumness, a measure of discord, can be defined in terms of the activation framework of Figure~\ref{setup}(a) \cite{acti,streltsov,quantumnessIJQI,taka,pianiadesso} as $Q_N(\chi_{A|B}) = \min_{U_B} N(\rho^{(U_B)}_{AB|M})$ (notice that the entanglement in the premeasurement states $\rho^{(U_B)}_{AB|M}$ is always distillable \cite{acti,streltsov}).
The trace-distance discord $D(\chi_{A|B})$ is defined as the minimum distance, in trace norm, between $\chi_{A|B}$ and the set of quantum-classical states $\varpi_{A|B} = \sum_n p_n \tau^n_A \otimes \ket{n}\bra{n}_B$, namely $D(\chi_{A|B})=\min_{\varpi_{A|B}}\|\chi_{A|B} - \varpi_{A|B}\|_1$ \cite{taka,sarandy}. 
If $B$ is a qubit, then $Q_N(\chi_{A|B})=D(\chi_{A|B})$ as first proven in \cite{taka}. We observe this equivalence experimentally in Figure~\ref{figattiva}(a).

{\bf Acknowledgments.}
G.A. acknowledges support from ESF, EPSRC, FQXi, and the Brazilian funding agency CAPES (Pesquisador Visitante Especial-Grant No. 108/2012). M.P. acknowledges support from NSERC, CIFAR, and Ontario Centres of Excellence. V.D., E.N. and F.S. acknowledge support by FIRB-Futuro in Ricerca HYTEQ and ERC (European Research Council) Starting Grant 3D-QUEST (3D-Quantum Integrated Optical Simulation; grant agreement no. 307783): http://www.3dquest.eu. We thank J.~Calsamiglia and A.~Streltsov for useful feedback on the Letter. G.A. thanks F.~G.~S.~L.~Brand\~{a}o, J.~Eisert, and A.~Winter for discussions, and D.~O.~Soares-Pinto at the Physics Institute of S\~{a}o Carlos for the kind hospitality during completion of this Letter.

\clearpage
\appendix

\numberwithin{equation}{section}
\makeatletter
\renewcommand{\thefigure}{S\@arabic\c@figure}

\begin{widetext}

\section*{SUPPLEMENTAL MATERIAL}

\section{Measures of correlations for the produced states}
The measures of discord and entanglement adopted in this paper are defined in the Methods section of the main manuscript. Here we provide additional details on their evaluation for the type of states we produced in our experiment.

For arbitrary Bell diagonal states $\chi_{A|B}$ of two qubits, the trace-distance discord $D$   is computable in closed form \cite{Ssarandy,Staka} and given by the intermediate singular value of the $3 \times 3$ correlation matrix $\tilde{\chi}_{A|B}$ of elements $(\tilde{\chi}_{A|B})_{i,j}={\rm Tr}\left[\chi_{A|B} ({\sigma_i}_A \otimes {\sigma_j}_B)\right]$, where $\sigma_{i,j=x,y,z}$ are the Pauli matrices. For the specific state in Eq.~(1) of the main text, $\tilde{\chi}^{(q)}_{A|B} = {\rm diag}(q,-q,2q-1)$ so theoretically ${D}_{\rm th}(\chi^{(q)}_{A|B})=q$.

When the state $\chi^{(q)}_{A|B}$ undergoes the activation protocol \cite{Sacti,Sstreltsov} as described in Figure~1 of the main article, or in other words undergoes an interaction with an ancilla qubit $M$ apt to probe subsystem $B$, if $U_B$ is parameterized as detailed in the Methods section of the main manuscript then the theoretical prediction for the entanglement, measured by the negativity \cite{Snegativity}, in the premeasurement state is
\begin{equation}
\label{negteo}
N_{\rm th}(\rho^{(q,\theta,\phi)}_{AB|M})=
\left\{
  \begin{array}{ll}
    \sqrt{\displaystyle\frac{(q-1) (3 q-1) \cos (4 \theta -8 \phi )+q (5 q-4)+1}{2}}, & \hbox{if $0\leq q<\frac13$;} \\
    q, & \hbox{if $\frac13\leq q \leq 1$.}
  \end{array}
\right.
\end{equation}
The above expression is periodic in the waveplates angles $\theta$ and $\phi$ with periodicity $\pi/2$ and $\pi/4$, respectively.
Minimizing Eq.~(\ref{negteo}) over $\theta,\phi$ straightforwardly gives
$\min_{(\theta,\phi)} N_{\rm th}(\rho^{(q,\theta,\phi)}_{AB|M})  \equiv {Q_N}_{\rm th}(\chi_{A|B}^{(q)}) = {D}_{\rm th}(\chi_{A|B}^{(q)}) = q$ as expected by definition, with the minimum attained by any measurement setting for $q \geq 1/3$, and by $\phi=\frac{\theta}{2}\pm\frac{\pi}{8}$, respectively, for $q<1/3$. Eq.~(\ref{negteo}) is plotted as smooth green translucent surface in the main Figure~2(b) as a function of $\theta$ and $\phi$ and closely resembles the net of measured datapoints obtained from state tomography.



\section{Continuity bounds for the negativity and construction of the net}

The activation theorem \cite{Sacti,Sstreltsov} predicts that if an input state $\chi_{A|B}$ has nonzero discord, then the premeasurement state $\rho_{AB|M}$ is entangled for any possible measurement settings. We aim to demonstrate this claim on the basis of a finite number of measurement settings.
Our goal is to show that for the whole continuum of measurement settings which we do not test, the premeasurement state $\rho_{AB|M}^{(\theta,\phi)}$ is guaranteed to remain entangled provided $\chi_{A|B}$ has nonzero discord. To this aim, we derive stringent continuity bounds for the negativity, that allow us to choose a net as loose and not experimentally demanding as possible, compatibly with the request above.

\subsection{First continuity bound}
We obtain the first continuity bound by using the following. Fix a state $\chi_{A|B}$, with $B$ a qubit. Consider a local projective measurement on the qubit $B$ performed in the orthonormal basis $\{\ket{n},\ket{n^\perp}\}$, the latter being uniquely described, up to irrelevant phase factors, by a Bloch vector {$\vec{n}$ (and its opposite $-\vec{n}$)}. The coherent interaction with the measurement apparatus $M$ gives rise to a premeasurement state $\rho^{\vec{n}}_{AB|M}$. Using the  conventions of this paper for the negativity, $N(\rho^{\vec{n}}_{AB|M}) = \|(\rho^{\vec{n}}_{AB|M})^\Gamma\|_1-1$, which amounts to a negativity normalized to unity for maximally entangled Bell states, it turns out that~\cite{Staka}
\begin{equation}
\label{eq:takaneg}
N(\rho^{\vec{n}}_{AB|M}) = 2\|\bra{n}\chi_{A|B}\ket{n^\perp}\|_1.
\end{equation}
We want now to bound the difference in negativity for measurements in directions $\vec{n}$ and $\vec{n}'$. We find:
\begin{equation}
\label{eq:bound1}
\begin{aligned}
N(\rho^{\vec{n}}_{AB|M})-N(\rho^{\vec{n}'}_{AB|M}) &\stackrel{(i)}{=} 2 (\|\bra{n}\chi_{A|B}\ket{n^\perp}\|_1 - \|\bra{n'}\chi_{A|B}\ket{n'^\perp}\|_1)\\
&\stackrel{(ii)}{=}2\Big(\| \tr_B\big(\ket{n^\perp}\bra{n}_B\chi_{A|B}\big)\|_1- \| \tr_B\big(\ket{n'^\perp}\bra{n'}_B\chi_{A|B}\big)\|_1\Big)\\
&\stackrel{(iii)}{=}2\Big(\| \tr_B\big(U_B\ket{n}\bra{n}_B\chi_{A|B}\big)\|_1- \| \tr_B\big(U_B\ket{n'}\bra{n'}_B\chi_{A|B}\big)\|_1\Big)\\
&\stackrel{(iv)}{\leq}2\| \tr_B\big(U_B(\ket{n}\bra{n}-\ket{n'}\bra{n'})_B\chi_{A|B}\big)\|_1\\
&\stackrel{(v)}{\leq}2\| U_B(\ket{n}\bra{n}-\ket{n'}\bra{n'})_B\chi_{A|B}\|_1\\
&\stackrel{(vi)}{=}2\| (\ket{n}\bra{n}-\ket{n'}\bra{n'})_B\chi_{A|B}\|_1\\
&\stackrel{(vii)}{\leq}\max_{\ket{\psi}_{A|B}}2\|\ \ket{n}\bra{n}-\ket{n'}\bra{n'})_B\proj{\psi}_{A|B}\ \|_1\\
&\stackrel{(viii)}{=}2\|\ \ket{n}\bra{n}-\ket{n'}\bra{n'}\ \|_\infty\\
&= 2 \sqrt{1- |\braket{n}{n'}|^2}\\
&=2 \sqrt{1- \frac{1+\vec{n}\cdot \vec{n}'}{2}}\\
&=\sqrt{2(1-\vec{n}\cdot\vec{n}')}.
\end{aligned}
\end{equation}
The steps above are justified as follows: $(i)$ by Eq.~\eqref{eq:takaneg}; $(ii)$ cyclic property of the trace; $(iii)$ existence of a $\pi$-rotation $U$ of the Bloch sphere around the $(\vec{n}\times \vec{n}')/ \|\vec{n}\times \vec{n}'\|$ axis that acts like the `NOT' in the plane containing $\ket{n},\ket{n'},\ket{n^\perp},\ket{n'^\perp}$ (even if a physical universal-NOT operation does not exist~\cite{Sunot,Sunot2}), mapping $\ket{n}$ into $\ket{n^\perp}$ and $\ket{n'}$ into $\ket{n'^\perp}$; $(iv)$ triangle inequality for the trace norm; $(v)$ contractivity of the trace norm under completely-positive trace-preserving maps (specifically, under a partial trace over $B$); $(vi)$ unitary invariance of the trace norm; $(vii)$ convexity of the trace norm; $(viii)$ by the fact that the operator norm of an operator $O$ is equal to $\|O\|_\infty = \max_{\ket{\psi}}\|O\ket{\psi}\|$, with the maximum taken over normalized states $\ket{\psi}$, together with the fact that the operator in the previous step is rank-$1$.

In our setting, let the Bloch vector of the measurement basis be parameterized by two angles, $\vec{n}=\vec{n}(\theta,\phi)$, e.g.~the angles of the waveplates as in our implementation. Suppose now that we have measured the negativity of the premeasurement state $N(\rho^{(\theta_j,\phi_k)}_{AB|M})$ for a discrete set of measurement settings, corresponding to a dataset of $m_j \times m_k$ phases $\{\big(\theta_j,\phi_k\big)\}_{j,k=1}^{m_{j,k}}$. Then the first continuity bound implies that for generic, unmeasured phases $(\theta,\phi)$, one has \begin{equation}\label{eq:estineg1}
N(\rho_{AB|M}^{(\theta,\phi)}) \geq N_{\rm low:1}(\rho_{AB|M}^{(\theta,\phi)}) \equiv \max_{(j,k)}
\left\{N(\rho_{AB|M}^{(\theta_j,\phi_k)}) - \sqrt{2[1-\vec{n}(\theta,\phi)\cdot\vec{n}(\theta_j,\phi_k)]} \right\}\,.
\end{equation}

\subsection{Second continuity bound}
A simpler continuity bound valid in general and not specialized to premeasurement states can be also derived as follows. For two generic bipartite states $\varrho$ and $\varsigma$, one has $|N(\varrho)-N(\varsigma)| \leq \|(\varrho-\varsigma)^\Gamma\|_1$,  as it is immediate to prove from the definition of negativity \cite{Snegativity} and the triangle inequality for the trace norm. In our setting, suppose again we have measured $N(\rho^{(q,\theta_j,\phi_k)}_{AB|M})$ in correspondence of a dataset of $m_j \times m_k$ phases $\{\big(\theta_j,\phi_k\big)\}_{j,k=1}^{m_{j,k}}$, then the continuity bound implies that for generic, unmeasured phases $(\theta,\phi)$, one has \begin{equation}\label{eq:estineg2}
N(\rho_{AB|M}^{(\theta,\phi)}) \geq  N_{\rm low:2}(\rho_{AB|M}^{(\theta,\phi)}) \equiv \max_{(j,k)} \left\{N(\rho_{AB|M}^{(\theta_j,\phi_k)})-\left\|\left[\rho_{AB|M}^{(\theta,\phi)}-\rho_{AB|M}^{(\theta_j,\phi_k)}\right]^\Gamma\right\|_1\right\}\,.\end{equation}
\subsection{Combined lower bound and measurement settings}
As stated, we are interested in a bound which stays nonzero to demonstrate the activation but  we want to limit the experimental effort. We want thus to determine a net of sampling points (bases for the measurement of qubit $B$) as sparse as possible to conclude that, in the presence of initial discord, entanglement across the $AB|M$ partition must always be created. The two bounds derived above serve this purpose. However, both of them rely on partial assumptions. The first bound relies on the assumption that both measured and unmeasured states are ideal premeasurement states. The derivation of second bound is valid for all states, and it is constructed from the tomographically reconstructed density matrices $\rho_{AB|M}^{(\theta_j,\phi_k)}$, but in its form~\eqref{eq:estineg2} it relies also on the specific structure of the expected output state $\rho_{AB|M}^{(\theta,\phi)}$ for unmeasured settings, that enters in the trace distance.  We  take a conservative view in order to make our conclusions about the necessary presence of entanglement as reliable as possible, and combine the two bounds by taking the {\it minimum} between them. Thus, our final lower bound $N_{\rm low}(\rho_{AB|M}^{(\theta,\phi)})$ on the output negativity of unmeasured states corresponding to (a continuum of) $U_B$ angles $(\theta,\phi)$, is
 \begin{equation}\label{eq:negbound}
N(\rho_{AB|M}^{(\theta,\phi)}) \geq  N_{\rm low}(\rho_{AB|M}^{(\theta,\phi)})=\min\left\{N_{\rm low:1}(\rho_{AB|M}^{(\theta,\phi)}),\,N_{\rm low:2}(\rho_{AB|M}^{(\theta,\phi)})\right\}\,.\end{equation}
We remark that this procedure is radically different from and more sensible (for our purposes) than a mere polynomial data interpolation.

We can now specialize to the initial states $\chi_{A|B}^{(q)}$ defined in Eq.~(1) of the main article, and to the corresponding premeasurement states $\rho_{AB|M}^{(q,\theta,\phi)})$, where we recall  from the Methods that in our implementation the Bloch vector defining the measurement basis reads\footnote{Notice that, due to the nontrivial trigonometric relation between the waveplates angles $\theta,\phi$ and the conventional polar and azimuthal angles on the Bloch sphere, there is not a one-to-one correspondence between individual measurement settings $(\theta_j,\phi_k)$, and individual Bloch vectors $\vec{n}(\theta_j,\phi_k)$; in particular, different pairs of choices of $(\theta_j,\phi_k)$ can lead to the same Bloch vector. The settings represented in Figure~2(a) of the main text and in Figure~\ref{figS1} correspond to the non-redundant values of $\pm\vec{n}(\theta_j,\phi_k)$.}
\begin{equation}\label{nbloch}
\vec{n}(\theta,\phi)=\{-\cos[2(\theta-2\phi)]\sin(2\theta),\,-\sin[2(\theta-2\phi)],\,\cos[2(\theta-2\phi)]\cos(2\theta)\}\,.
\end{equation}

We performed an {\it a priori} numerical simulation based on \eqref{eq:negbound} and on the theoretical expectation for the negativity, given by Eq.~(\ref{negteo}), to investigate the largest possible spacing we could select for variations of $(\theta,\phi)$, in order to ensure that the lower bound $N_{\rm low}(\rho_{AB|M}^{(q,\theta,\phi)})$ stayed nonnegative in the relevant parameter range. As a result we found that, to prove conclusively activation, it is enough to take $\theta_j$ and $\phi_k$ varying in their interval of interest $\theta \in [0,\pi/2], \phi \in [0,\pi/4]$ by a step of $\pi/12$, which we have adopted to define the measurement settings that have been automatized for the data acquisition of this experiment. Figure~2(b) of the main article confirms that this is a {satisfactory choice}. Notice that, with these settings, the corresponding measurement basis vectors $\pm\vec{n}(\theta_j,\phi_k)$ are not equispaced on the Bloch sphere of qubit $B$ (see Figure~\ref{figS1}). This is intentional: we choose in fact to have a denser net around the poles of the $y$ axis, because the negativity of our premeasurement states (for $q<1/3$) is minimized in those regions, as evident from Eq.~(\ref{negteo}) and from Figure~2(a) of the main article; in this way, having more points in such a region, we avoid that the lower bound on $N$ falls below zero for small but nonzero values of $q$.

In the next section we provide more rigorous arguments in support of our net construction, using notions borrowed from the geometry of metric spaces.

\subsection{$\boldsymbol{\epsilon}$-net on the Bloch sphere}

Our construction is reminiscent of a so-called $\epsilon$-net~\cite{Senet1}. From our perspective, an $\epsilon$-net on a metric space ${\cal M}$ can be roughly described as an optimal sampling of the space. An $\epsilon$-net is a subset $X$ of points of ${\cal M}$ such that any point in ${\cal M}$ is not more than $\epsilon$-away from a point in $X$, i.e.~the net ``covers'' the space ${\cal M}$, and such that the elements of the net are at least $\epsilon$-away from each other. The crucial {aspect} is that such a subset $X$ can be discrete, while the original space ${\cal M}$ is continuous. For instance, a discrete set $X=\{x_i\}_{i=1}^n$ of $n$ points realizes an $\epsilon$-net for the metric space ${\cal M}$ if the union of  $n$  ``balls'' of radius $\epsilon$ centered at points $\{x_i\}$ covers ${\cal M}$, while the $n$ ``balls'' centered at the same points but with radius $\epsilon/2$ are all disjoint.

Our choice of the measurement settings $(\theta_j,\phi_k$) obviously defines an $\epsilon$-net for the two-dimensional space spanned by the waveplates angles $(\theta,\phi)$ equipped with the Euclidean distance, for any $\frac{\pi}{12 \sqrt{2}}  \leq \epsilon< \frac{\pi}{12}$. What is more interesting, though, is to investigate whether the corresponding basis vectors $\vec{n}(\theta_j,\phi_k)$ also define a proper $\epsilon$-net on the Bloch sphere of qubit $B$. To proceed, we need to adopt a metric on the unit sphere $SO(3)$.

\begin{figure*}[t]
\includegraphics[width=17cm]{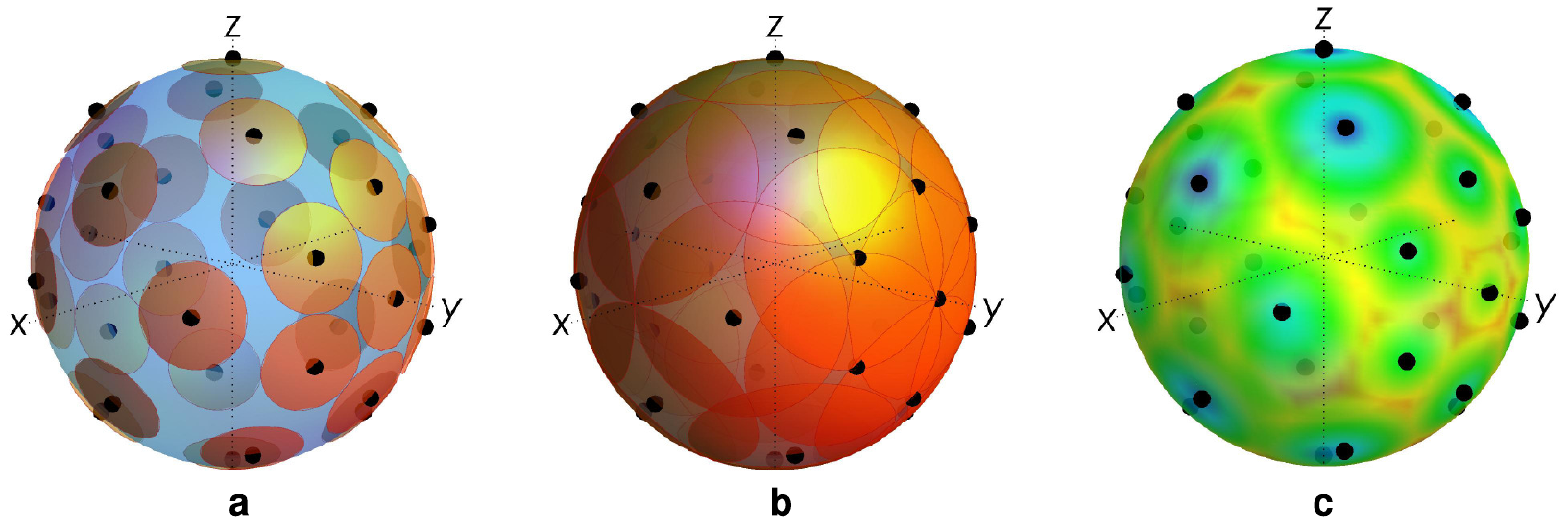}
\caption{{\bf $\boldsymbol{\epsilon}$-nets.} Our choice of a finite number of measurement bases, parameterized  by the Bloch vectors $\pm \vec{n}(\theta_j,\phi_k)$ of Eq.~(\ref{nbloch}) with waveplates angles $\theta_j = j \frac{\pi}{12}, \phi_k=k \frac{\pi}{12}$, $j=0,\ldots,6$, $k=0,\ldots,3$, realize an $\epsilon$-net for the Bloch sphere of qubit $B$ equipped with the Euclidean distance, Eq.~(\ref{eq:euc}), for $\epsilon=1/2$. (a) Spherical caps centered at $\pm \vec{n}(\theta_j,\phi_k)$ with base radius $a_{\epsilon/2}$ are disjoint (a packing set). (b) Spherical caps centered at $\pm \vec{n}(\theta_j,\phi_k)$ with base radius $a_{\epsilon}$ patch the whole sphere (a covering set).   (c) Density plot of the first lower bound $N_{\rm low:1}(\rho_{AB|M}^{(q)})$ on the negativity of the premeasurement state $\rho_{AB|M}^{(q)}$ evaluated from Eq.~(\ref{eq:bound1}) using the experimental data for $q=0.2$; notice the geometric nature of the bound, which is manifest in the distribution of its contours around the measured points, which matches that of the spherical caps constructed from the $\epsilon$-net. The color in the density plot varies from red to blue corresponding to $N_{\rm low:1}$ ranging from $0$ to $1$ (the actual minimum reached in the plot is $N_{\rm low:1} \approx 0.1$ while the maximum  is $N_{\rm low:1} \approx 0.6$).
}
\label{figS1}
\end{figure*}

We observe that, remarkably, the first continuity bound on the negativity of premeasurement states, derived in Eq.~(\ref{eq:bound1}), coincides exactly with the Euclidean metric between two points on the Bloch sphere associated to Bloch vectors $\vec{n}$, $\vec{n}'$, respectively, i.e., with
\begin{equation}\label{eq:euc}
\|\vec{n}-\vec{n}'\|_{\rm E}=\sqrt{(\vec{n}-\vec{n}')\cdot(\vec{n}-\vec{n}')} = \sqrt{2(1-\vec{n}\cdot\vec{n}')}\,.
\end{equation}
We are then naturally induced to choose this metric (that is different from other valid choices such as the Riemannian metric on the surface of the sphere) for our analysis. Thus, we are interested in covering the Bloch sphere by spherical caps centered at our chosen basis vectors $\vec{n}(\theta_j,\phi_k)$ and characterized by base circles with radius $a_\epsilon$, such that the vectors $\vec{n}'$ forming the base circle of each cap are at distance $\epsilon$ from the center, namely $\|\vec{n}(\theta_j,\phi_k)-\vec{n}'\|_{\rm E} =\epsilon$. From elementary geometry we see that the cap base radius relates to $\epsilon$ according to the formula
$$a_\epsilon = \frac14 \sqrt{\epsilon^2(4-\epsilon^2)}.$$
In Figure~\ref{figS1}, we show that by choosing $\epsilon=1/2$ our measurement settings realize precisely an $\epsilon$-net for the Bloch sphere.

The problem of defining sampling sets such that the output negativity stays nonzero, according to the lower bound of Eq.~(\ref{eq:bound1}), can then be interpreted geometrically in terms of defining minimal $\epsilon$-nets for the Bloch sphere. In Figure~\ref{figS1}(c) we overlay the Bloch sphere with a density plot of the lower bound $N_{\rm low:1}(\rho_{AB|M}^{(q)})$ on the output negativity, constructed from the experimental data as in Figure~2 of the main article, for $q=0.2$. This state has small but nonzero input discord, and we can certify that its negativity stays nonzero for all possible Bloch measurement bases. Notice how the points when the bound reaches its minimum (staying nonzero) are those which maximize the Euclidean distance from all the neighboring measured positions. At these points, such as the orange region in the top-left corner, or the rim around the poles on the $y$ axis, the Bloch sphere would not be covered anymore if one decreased the value of $\epsilon$.

\section{Experimental details}

\subsection{State tomography}

An example of tomographically reconstructed state is reported in Fig.~\ref{figstomo}.

\begin{figure}[hb!]
\includegraphics[width=11cm]{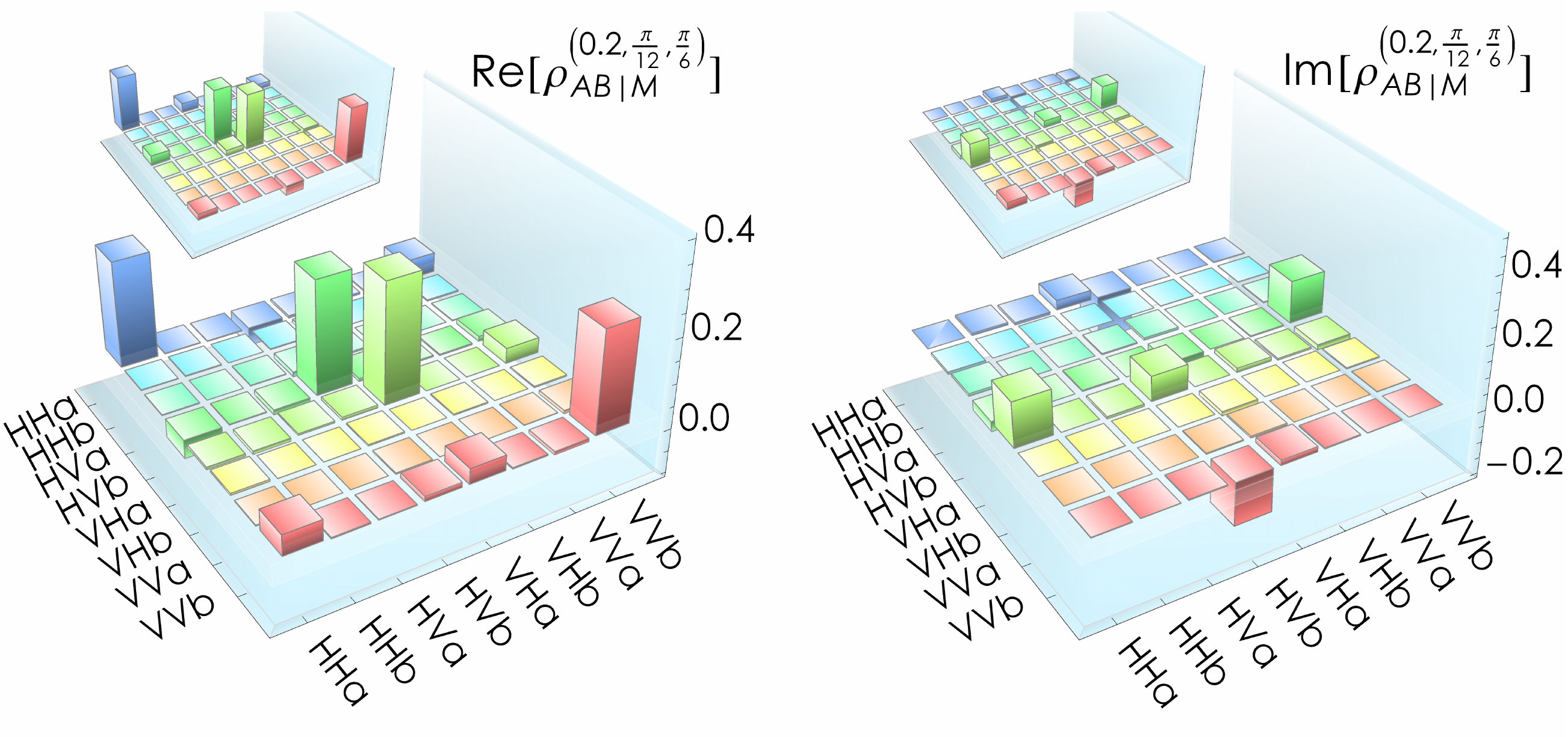}
\caption{An example of tomographically reconstructed premeasurement state $\rho^{(q,\theta,\phi)}_{AB|M}$ of the polarization qubits $A$, $B$, and the path qubit $M$ is reported for $q=0.2$, $\theta=\frac{\pi}{12}$, $\phi=\frac{\pi}{6}$. The insets display the theoretical prediction for a comparison; the fidelity between the experimentally reconstructed density matrix and the theoretical one was $90.1\%$.}
\label{figstomo}
\end{figure}

In Table I we report all the experimental settings for the waveplates in the Mach-Zehnder interferometer (experimentally implemented as a Sagnac interferometer) needed to measure polarization and path of photon 2 (Figure~1 in the main article). For each set of four orthogonal states we report the waveplates angles and the corresponding detector of each state.

\begin{table*}[h]
{\small
\begin{tabular}{|c|c|c|c|c|c|c|c|c|c|c|}
\hline
\hline
\textbf{Set} &\textbf{State} & \textbf{QWPa} & \textbf{HWPa} & \textbf{QWPb} & \textbf{HWPb} & \textbf{QWPc} & \textbf{HWPc} & \textbf{QWPd} & \textbf{HWPd} & \textbf{DB}\\
\hline\hline
\multirow{4}{*}{I}&$\ket{H,a}$ &\multirow{4}{*}{0}&\multirow{4}{*}{0}&\multirow{4}{*}{0}&\multirow{4}{*}{45}&\multirow{4}{*}{0}&\multirow{4}{*}{0}&\multirow{4}{*}{0}&\multirow{4}{*}{0}&1\\
 &$\ket{H,b}$ &  & & & & & & & &2\\
 &$\ket{V,a}$ & & & & & & & & &3\\
 &$\ket{V,b}$ &  & & & & & & & &4\\ \hline
\multirow{4}{*}{II} &$\ket{+,a}$&\multirow{4}{*}{45}&\multirow{4}{*}{22.5}&\multirow{4}{*}{45}&\multirow{4}{*}{-22.5}&\multirow{4}{*}{0}&\multirow{4}{*}{0}&\multirow{4}{*}{0}&\multirow{4}{*}{0}&1\\
 &$\ket{+,b}$ & & & & & & & & &2\\
 &$\ket{-,a}$ & & & & & & & & &3\\
 &$\ket{-,b}$ && & & & & & & &4\\ \hline
 \multirow{4}{*}{III} &$\ket{R,a}$&\multirow{4}{*}{0}&\multirow{4}{*}{22.5}&\multirow{4}{*}{0}&\multirow{4}{*}{-22.5}&\multirow{4}{*}{0}&\multirow{4}{*}{0}&\multirow{4}{*}{0}&\multirow{4}{*}{0}&1\\
 &$\ket{R,b}$ & & & & & & & & &2\\
 &$\ket{L,a}$ & & & & & & & & &3\\
 &$\ket{L,b}$ && & & & & & & &4\\ \hline
\multirow{4}{*}{IV} &$\ket{H,+}$&\multirow{4}{*}{0}&\multirow{4}{*}{0}&\multirow{4}{*}{0}&\multirow{4}{*}{45}&\multirow{4}{*}{45}&\multirow{4}{*}{22.5}&\multirow{4}{*}{45}&\multirow{4}{*}{22.5}&1\\
 &$\ket{H,-}$ & & & & & & & & &2\\
 &$\ket{V,+}$ & & & & & & & & &4\\
 &$\ket{V,-}$ & & & & & & & & &3\\ \hline
\multirow{4}{*}{V} &$\ket{+,+}$&\multirow{4}{*}{45}&\multirow{4}{*}{22.5}&\multirow{4}{*}{45}&\multirow{4}{*}{-22.5}&\multirow{4}{*}{45}&\multirow{4}{*}{22.5}&\multirow{4}{*}{45}&\multirow{4}{*}{22.5}&1\\
 &$\ket{+,-}$ & & & & & & & & &2\\
 &$\ket{-,+}$& & & & & & & & &4\\
 &$\ket{-,-}$ & & & & & & & & &3\\ \hline
 \multirow{4}{*}{VI} &$\ket{R,+}$&\multirow{4}{*}{0}&\multirow{4}{*}{22.5}&\multirow{4}{*}{0}&\multirow{4}{*}{-22.5}&\multirow{4}{*}{45}&\multirow{4}{*}{22.5}&\multirow{4}{*}{45}&\multirow{4}{*}{22.5}&1\\
 &$\ket{R,-}$ & & & & & & & & &2\\
 &$\ket{L,+}$& & & & & & & & &4\\
 &$\ket{L,-}$ & & & & & & & & &3\\ \hline
 \multirow{4}{*}{VII}&$\ket{H,r}$ &\multirow{4}{*}{0}&\multirow{4}{*}{0}&\multirow{4}{*}{0}&\multirow{4}{*}{45}&\multirow{4}{*}{0}&\multirow{4}{*}{22.5}&\multirow{4}{*}{0}&\multirow{4}{*}{22.5}&1\\
 &$\ket{H,l}$ &  & & & & & & & &2\\
 &$\ket{V,r}$ & & & & & & & & &3\\
 &$\ket{V,l}$ &  & & & & & & & &4\\ \hline
 \multirow{4}{*}{VIII} &$\ket{+,r}$&\multirow{4}{*}{45}&\multirow{4}{*}{22.5}&\multirow{4}{*}{45}&\multirow{4}{*}{-22.5}&\multirow{4}{*}{0}&\multirow{4}{*}{22.5}&\multirow{4}{*}{0}&\multirow{4}{*}{22.5}&1\\
 &$\ket{+,l}$ & & & & & & & & &2\\
 &$\ket{-,r}$& & & & & & & & &4\\
 &$\ket{-,l}$ & & & & & & & & &3\\ \hline
  \multirow{4}{*}{IX} &$\ket{R,r}$&\multirow{4}{*}{0}&\multirow{4}{*}{22.5}&\multirow{4}{*}{0}&\multirow{4}{*}{-22.5}&\multirow{4}{*}{0}&\multirow{4}{*}{22.5}&\multirow{4}{*}{0}&\multirow{4}{*}{22.5}&1\\
 &$\ket{R,l}$ & & & & & & & & &2\\
 &$\ket{L,r}$& & & & & & & & &4\\
 &$\ket{L,l}$ & & & & & & & & &3\\ \hline
\hline
\end{tabular}
}
\caption{Settings for the polarization and path measurements (qubit $B$ and $M$) of photon BM. }
\label{prob}
\end{table*}

In the two arms of the interferometer QWPa and HWPa are respectively
the first and second waveplates of WPa while QWPb and HWPb are first
and second waveplates of WPb. After the interferometer QWPc and HWPc
are part of the analysis stage ASc while QWPd and HWPd belong to ASd.

\subsection{Quantitative demonstration of the activation protocol}
Here we detail the experimental procedures to obtain the results of Figure~3(a) of the main text.
To demonstrate that input bipartite discord activates quantitatively into bipartite entanglement with the apparatus \cite{Sacti,Sstreltsov}, we have adopted two different procedures for the evaluation of  $D(\chi_{A|B}^{(q)})$ (input discord) and $\min_{(\theta,\phi)} N(\rho_{AB|M}^{(q,\theta,\phi)}$ (minimum output bipartite entanglement) from the measured data. For the input trace-distance discord $D(\chi_{A|B}^{(q)})$, we have reconstructed the density matrix of qubits $A$ and $B$ after $U_B$ but removing the C-NOT operation; we denote the reconstructed state by $\chi_{A|B}^{(q,\theta_j,\phi_k)}$ for each of the $28$ measurement settings $(j,k)$. Theoretically, $D(\chi_{A|B}^{(q,\theta_j,\phi_k)})$ should be independent of $j,k$ since any setting just amounts to a local unitary which does not change correlations by construction. To account for possible experimental variations, we take mean and standard deviation of the array $\left\{D(\chi_{A|B}^{(q,\theta_j,\phi_k)})\right\}_{(j,k)}$ over $j=0,\ldots,6,\,k=0,\ldots,3$ as the experimental value for $D(\chi_{A|B}^{(q)})$ and its error, respectively. Here each $D(\chi_{A|B}^{(q,\theta_j,\phi_k)})$ is calculated from the reconstructed states by numerically minimizing the trace distance from the set of quantum-classical states using an algorithm analogous to the one described in \cite{Scicca}.\footnote{The results are in agreement, within the reported error bars in Fig. 3(a), with the closed formula for $D$ available from \cite{Ssarandy,Staka} evaluated approximating the tomographically reconstructed input states with Bell diagonal states.} These data are plotted as orange filled squares. Concerning the minimum output entanglement in the premeasurement state, we take the minimum of the array of measured data for the negativity $N(\rho_{AB|M}^{(q,\theta_j,\phi_k)})$ as the experimental value for each $q$, with its error given by Poissonian statistics of photon events.

\subsection{Entanglement witnesses}
Here we provide details on  the measurement of two entanglement witnesses, reported in Figure~3(b) of the main text. Specifically, $W^{(2)}_{A|B} = \mathbb{I}_{A|B}/2 - (\ket{\Phi^-}\bra{\Phi^-}_{A|B})^\Gamma$ is a witness for bipartite entanglement in the initial state $\chi^{(q)}_{A|B}$, while $W^{(3)}_{A|B|M} = \mathbb{I}_{A|B}/2 - (\ket{\widetilde{\rm GHZ}}\bra{\widetilde{\rm GHZ}}_{A|B|M})$ is a witness for genuine tripartite GHZ--type entanglement among $A$, $B$ and $M$ in the premeasurement state $\rho^{(q)}_{AB|M}$, where   $\ket{\widetilde{\rm GHZ}} = \frac12 \left(-\ket{HHa}-i \ket{HVb}+i\ket{VHa}+\ket{VVb}\right)$ is a locally rotated three-qubit GHZ state. Entanglement witnesses are usually state-dependent and a single linear witness can only detect entanglement for some particular subclass of states. For this analysis, we fix therefore the measurement setting to $(\theta_j=\frac{\pi}{4},\phi_k=0)$, as from the investigation in Figure~2(b) of the main text we have learned that this is a worst case scenario which minimizes the created negativity.

Each witness can be decomposed in terms of projections which can be implemented on the appropriate degrees of freedom of photons A and BM according to the procedure already used for tomography. Explicitly,
$$W^{(2)}_{A|B}=(\mathbb{I}\otimes \mathbb{I}-\sigma_x\otimes \sigma_x-\sigma_y\otimes \sigma_y+\sigma_z\otimes \sigma_z)/4\,,\quad \mbox{and}$$  $$W^{(3)}_{A|B}= (3 \mathbb{I}\otimes \mathbb{I}\otimes \mathbb{I}-\mathbb{I}\otimes \sigma_z\otimes \sigma_z+\sigma_x\otimes \sigma_x\otimes \sigma_x-\sigma_x\otimes \sigma_y\otimes \sigma_y+\sigma_y\otimes \mathbb{I}\otimes \sigma_z+\sigma_y\otimes \sigma_z\otimes \mathbb{I}-\sigma_z\otimes \sigma_x\otimes \sigma_y-\sigma_z\otimes \sigma _y\otimes \sigma _x)/8\,,$$ where $\{\sigma_{x,y,z}\}$ are Pauli matrices.

The measured expectation values of $W^{(2)}$ on the input state $\chi_{A|B}^{(q)}$ and of  $W^{(3)}$ on the output premeasurement state $\rho^{(q,\frac{\pi}{4},0)}_{AB|M}$ are plotted in Figure~3(b) of the main text using orange filled squares and blue empty circles, respectively.


\clearpage
\end{widetext}


\begin{thebibliography}{99}


\bibitem{NC} M. A. Nielsen and I. L. Chuang, "Quantum computation and quantum information", Cambridge University Press (2010).


\bibitem{entanglement}  R. Horodecki, P. Horodecki, M. Horodecki, and K. Horodecki,  {Rev. Mod. Phys.} {\bf 81}, 865 (2009).


\bibitem{vonneumann} J. von Neumann, {\it Mathematical Foundation of Quantum Mechanics} (Princeton University Press, Princeton, NJ, 1955).



\bibitem{zurekrev} W. H. Zurek, {
Rev. Mod. Phys.} {\bf 75}, 715 (2003).


\bibitem{zurek} H. Ollivier and W. H. Zurek,  {Phys. Rev. Lett.} {\bf 88}, 017901
(2001).

\bibitem{vedral} L. Henderson and V. Vedral, {J. Phys. A.: Math. Gen.} {\bf 34}, 6899 (2001).


\bibitem{reviewmodi}
K. Modi, A. Brodutch, H. Cable, T. Paterek, and V. Vedral, {Rev. Mod. Phys.} {\bf 84}, 1655 (2012).


\bibitem{acti} M. Piani, S. Gharibian, G. Adesso, J. Calsamiglia, P. Horodecki, and A. Winter, {Phys. Rev. Lett.} {\bf 106}, 220403 (2011).

\bibitem{streltsov} A. Streltsov, H. Kampermann, and D. Bruss, {Phys. Rev. Lett.} {\bf 106}, 160401 (2011).

\bibitem{lqu}
D. Girolami, T. Tufarelli, and G. Adesso, {Phys. Rev. Lett.} {\bf 110}, 240402 (2013).

\bibitem{newalex}
A. Streltsov, and W. H. Zurek, {Phys. Rev. Lett.} {\bf 111}, 040401  (2013).

\bibitem{marcodarwin}
F. G. S. L. Brand\~{a}o, M. Piani, and P. Horodecki,
arXiv:1310.8640.


\bibitem{datta}
A. Datta, A. Shaji, and C. M. Caves,  {Phys. Rev. Lett.} {\bf 100}, 050502 (2008).

\bibitem{white}
B. P. Lanyon, M. Barbieri, M. P. Almeida, and A. G. White,  {Phys. Rev. Lett.} {\bf 101}, 200501 (2008).

\bibitem{concordant}
B. Eastin, arXiv:1006.4402 (2013).

\bibitem{melo} R. Chaves and F. de Melo, {Phys. Rev. A} {\bf 84}, 022324 (2011).


\bibitem{cavalcantidiscord} D. Cavalcanti, L. Aolita, S. Boixo, K. Modi, M. Piani, and A. Winter, {Phys. Rev. A} {\bf 83}, 032324 (2011).

\bibitem{dattaoper}
V. Madhok and A. Datta, {Phys. Rev. A} {\bf 83}, 032323 (2011).

\bibitem{npgu}
M. Gu  {\it et al.}, {Nature Phys.} {\bf 8}, 671 (2012).

\bibitem{npdakic}
B. Dakic, {\it et al.},  {Nature Phys.} {\bf 8}, 666 (2012).

\bibitem{entdist1}
A. Streltsov, H. Kampermann, and D. Bruss, {Phys. Rev. Lett.} {\bf 108}, 250501 (2012).

\bibitem{entdist2}
T. K. Chuan, J. Maillard, K. Modi, T. Paterek, M. Paternostro, and M. Piani, {Phys. Rev. Lett.} {\bf 109}, 070501 (2012).

\bibitem{entdistexp}
A. Fedrizzi {\it et al.}, {Phys. Rev. Lett.} {\bf 111}, 230504 (2013).

\bibitem{interpower}
D. Girolami {\it et al.}, 	arXiv:1309.1472 (2013).

\bibitem{merali}
Z. Merali, {Nature} {\bf 474}, 24 (2011).




\bibitem{horo} P. Horodecki, J. Tuziemski, P. Mazurek, and R. Horodecki,  arXiv:1306.4938 (2013).


\bibitem{renzie}
G. L. Giorgi, 	Phys. Rev. A {\bf 88}, 022315 (2013)

\bibitem{maurolaura}
L. Mazzola and M. Paternostro, {Sci. Rep.} {\bf 1}, 199 (2011).

\bibitem{giovannetti}
A. Farace, F. Ciccarello, R. Fazio, and V. Giovannetti, Phys. Rev. A {\bf 89}, 022335 (2014).

\bibitem{quantumnessIJQI} S. Gharibian, M. Piani, G. Adesso, J. Calsamiglia, and P. Horodecki, {Int. J. Quant. Inf.} {\bf 9}, 1701 (2011).

\bibitem{coleshierarchy} P. Coles, {Phys. Rev. A} {\bf 86}, 062334 (2012).

\bibitem{modiunif} K. Modi, T. Paterek, W. Son, V. Vedral, and  M. Williamson,
{Phys. Rev. Lett.} {\bf 104}, 080501 (2010).

\bibitem{taka}
T. Nakano, M. Piani, and G. Adesso,
{Phys. Rev. A} {\bf 88}, 012117 (2013).


\bibitem{enet1}
W. A. Sutherland,  {\it Introduction to metric and topological spaces} (Oxford University Press, Oxford, 1975).

\bibitem{enet2}
P. Hayden, D. Leung, P. W. Shor, and A. Winter,
{Commun. Math. Phys.} {\bf 250}, 371 (2004).


\bibitem{twirling} J. Emerson, {\it et al.}, {Science} {\bf 317}, 1893 (2007).

\bibitem{lobino2008} M. Lobino, {\it et al.}, {Science} {\bf 322}, 563 (2008).

\bibitem{pianiadesso}
M. Piani and G.  Adesso,  {Phys. Rev. A} {\bf 85}, 040301(R) (2012).

\bibitem{witness}
O. G\"uhne and G. Toth, {Phys. Rep.} {\bf 474}, 1 (2009).

\bibitem{Kwia95}
P. Kwiat, K. Mattle, H. Weinfurter, and A. Zeilinger,
 Phys. Rev. Lett. \textbf{75}, 4337 (1995).

\bibitem{marcobound}
J. Lavoie, R. Kaltenbaek, M. Piani, and K. J. Resch, {Phys. Rev. Lett.} {\bf 105}, 130501 (2010)

\bibitem{Jame01} D. F. V. James, P. G.  Kwiat, W. J.   Munro,  and A. G. White,
{Phys. Rev. A} {\bf 64}, 052312 (2001).


\bibitem{Naga12} E. Nagali, S. Felicetti, P.-L. de Assis, V. D'Ambrosio, R. Filip, and F. Sciarrino,  {Sci. Rep.} {\bf 2}, 443 (2012).

\bibitem{Walb06} S. P. Walborn, P. H. Souto Ribeiro, L.  Davidovich, F. Mintert, and A. Buchleitner,  {Nature} {\bf 440}, 1022 (2006).

\bibitem{Okam11} R. Okamoto,  J. L.  O'Brien, H. F.  Hofmann, and S. Takeuchi, {Proc. Nat. Acad. Sci.} {\bf 108}, 10067 (2011).

\bibitem{epaps} See Supplemental Material for additional details on the experimental and theoretical procedures; this includes Refs.~\cite{unot,unot2,cicca}.

\bibitem{unot} V. Buzek, M. Hillery, and R. F. Werner, {Phys. Rev. A} {\bf 60}, 2626(R) (1999).

\bibitem{unot2}
F. De Martini, V. Buzek, F. Sciarrino, and C. Sias,
{Nature} {\bf 419}, 815 (2002).


\bibitem{cicca}
F. Ciccarello, T. Tufarelli, and V. Giovannetti, {New J. Phys.} {\bf 16}, 013038 (2014).


\bibitem{negativity} G. Vidal and R. F. Werner, {Phys. Rev. A} {\bf 65}, 032314 (2002).


\bibitem{isabela} I. A. Silva  {\it et al.}, {Phys. Rev. Lett.} {\bf 110}, 140501 (2013).

\bibitem{sarandy}
F. M. Paula, T. R.  de Oliveira, and M. S.  Sarandy,   {Phys. Rev. A} {\bf 87},
 064101 (2013).

\bibitem{noteU}
The operator $U_B$ is decomposed as a sequence of two rotated waveplates, $U_B= R^T(\phi)U_B^H R(\phi) R^T(\theta) U_B^Q R(\theta)$, where $U_B^H={\rm diag}(1,-1)$ and $U_B^Q={\rm diag}(1,i)$ correspond to a half- and a quarter-waveplate respectively (with optical axes parallel to the horizontal polarization) and $R(\alpha)={{\cos\alpha\,-\sin\alpha}\choose{\sin\alpha\,\,\cos\alpha}}$ is a rotation. The measurement interaction resulting from applying a C-NOT gate after $U_B$ is equivalent (up to local unitaries) to the one achieving $V_{BM} \ket{n}_B \ket{a}_M = \ket{n}_B\ket{n}_M$ where $\ket{n}_B = U_B^\dagger\ket{H}_B$ is a state with Bloch vector $\vec{n}(\theta,\phi)=\{-\cos[2(\theta-2\phi)]\sin(2\theta),\,-\sin[2(\theta-2\phi)],\,\cos[2(\theta-2\phi)]\cos(2\theta)\}$. Correspondingly, $U_B^\dagger\ket{V}_B$ has Bloch vector $-\vec{n}(\theta,\phi)$. By periodicity, we can restrict to a parameter range $\theta \in [0,\pi/2], \phi \in [0,\pi/4]$, so that the basis vectors $\pm\vec{n}(\theta,\phi)$ cover the whole Bloch sphere of qubit $B$.

\bibitem{Geremia} A. Politi, M. J. Cryan, J. G. Rarity, S. Y. Yu,  and J. L. O'Brien, {Science} {\bf 320}, 646  (2008).

\bibitem{Cres11} A. Crespi,  {\it et al.}, {Nature Commun.} {\bf 2}, 566 (2011).

\bibitem{Cres13} A. Crespi, {\it et al.}, {Nature Photon.} {\bf 7}, 545 (2013).

\end{thebibliography}

\begin{thebibliography}{99}



\bibitem[S1]{Ssarandy}
Paula, F. M., de Oliveira, T. R. \& Sarandy, M. S. Geometric quantum discord through the Schatten 1-norm.  {\it Phys. Rev. A} {\bf 87},
 064101 (2013).

\bibitem[S2]{Staka}
Nakano, T., Piani, M. \& Adesso, G.
Negativity of quantumness and its interpretations. {\it Phys. Rev. A} {\bf 88}, 012117 (2013).



\bibitem[S3]{Snegativity} Vidal, G. and  Werner, R. F.  Computable measure of entanglement, {\it Phys. Rev. A} {\bf 65}, 032314 (2002).


\bibitem[S4]{Sacti} Piani, M. Gharibian, S. Adesso, G., Calsamiglia, J., Horodecki, P. \& Winter, A. All nonclassical correlations can be activated into distillable entanglement. {\it Phys. Rev. Lett.} {\bf 106}, 220403 (2011).

\bibitem[S5]{Sstreltsov} Streltsov, A., Kampermann, H. \& Bruss, D. Linking quantum discord to entanglement in a measurement, {\it  Phys. Rev. Lett.} {\bf 106}, 160401 (2011).



\bibitem[S6]{Spianiadesso}
Piani, M. \& Adesso, G. Quantumness of correlations revealed in local measurements exceeds entanglement. {\it Phys. Rev. A} {\bf 85}, 040301(R) (2012).

\bibitem[S7]{Sunot} Buzek, V., Hillery, M. \& Werner, R. F. Optimal manipulations with qubits: universal-NOT gate. {\it Phys. Rev. A} {\bf 60}, 2626(R) (1999).

\bibitem[S8]{Sunot2}
De Martini, F., Buzek, V., Sciarrino, F. \& Sias, C.
Experimental realization of the quantum universal NOT gate.
{\it Nature} {\bf 419}, 815 (2002).


\bibitem[S9]{Senet1}
Sutherland, W. A. {\it Introduction to metric and topological spaces} (Oxford University Press, Oxford, 1975).

\bibitem[S10]{Scicca}
Ciccarello, F., Tufarelli, T., Giovannetti, V. Towards computability of trace distance discord. {\it New J. Phys.} {\bf 16}, 013038 (2014).

\end{thebibliography}
\end{document}